\documentclass[aps,showpacs,pre,preprint]{revtex4}
\usepackage{graphicx,psfrag,amsmath,amssymb,amsfonts}

\begin{document}
\title{Statistical Mechanics of Lam\'e Solitons}

\author{{Ioana Bena\footnote{Ioana.Bena@physics.unige.ch}}}
\affiliation{Department of Theoretical Physics, 
University of Geneva, CH-1211 Geneva 4, Switzerland}
\author{Avinash Khare\footnote{khare@iopb.res.in}}
\affiliation{Institute of Physics, Bhubaneswar, Orissa 751005, India}
\author{Avadh Saxena\footnote{abs@viking.lanl.gov}}
\affiliation{Theoretical Division, Los Alamos National Laboratory, 
Los Alamos, 
New Mexico 87545, USA}

\date{\today}

\begin{abstract}
We study the exact statistical mechanics of Lam\'e solitons 
using a transfer matrix method. This requires a knowledge
of the first forbidden band of the corresponding 
Schr\"odinger equation with the periodic Lam\'e potential.  Since 
the latter is a quasi-exactly solvable system, an analytical evaluation
of the partition function can be done only for a few temperatures.  We also 
study approximately the finite temperature thermodynamics using the 
ideal kink gas phenomenology. The zero-temperature 
``thermodynamics" of the soliton lattice solutions is also addressed. 
Moreover, in appropriate limits our results 
reduce to that of the sine-Gordon problem. 
\end{abstract}
\maketitle

\newpage

\section*{1. Introduction}

Systems described by one-dimensional nonlinear scalar fields and  
governed by nonlinear energy functionals are ubiquitous in physics. 
Frequently the governing dynamical equations of these fields admit 
large-amplitude, localized, robust, particle-like solutions 
-- the solitary waves~\cite{sg}. In anharmonic models of various 
condensed-matter systems~\cite{bishop,currie} the presence 
of these nonlinear excitations is thermally controlled. It is therefore 
important to proceed to a statistical mechanics investigation 
of the system,  considering the different excitations that 
are potentially present, as well as their interactions, since they
contribute nontrivially to the thermodynamical properties.

Over the years, extensive research has been carried 
out seeking the soliton solutions of  the integrable 
sine-Gordon (sG) equation~\cite{sg} and its 
(non-integrable)  counterpart,  
the double sine-Gordon equation (DsG)~\cite{leung,peyrard,saxena}. 
The statistical and thermodynamical properties 
for both the sG and  DsG potentials were studied~\cite{gupta,deleonardis}. 

There have also been advances in the study of the {\em hyperbolic} 
analogues of these problems, the sine-hyperbolic Gordon (shG) equation~\cite{shg}, respectively  the  double sine-hyperbolic Gordon (DshG) equation \cite{dshg} (which is non-integrable).
The statistical mechanics of double sinh-Gordon solitons was also addressed in~\cite{dshg}.

However,  the study of the {\em elliptic} analogues of the sG and DsG problems 
is less advanced.
The elliptic generalization of the sG is the {\em Lam\'e equation} ~\cite{arscott} and its single soliton and soliton lattice solutions have  been worked out only recently~\cite{bena}.  
One would surmise that the elliptic generalization  of the DsG equation is 
the Associated Lam\'e  equation~\cite{magnus,khare,ganguly}.
Nevertheless, we found~ \cite{bena} that this is not the case, 
and this generalization is still an open problem.

There are many physical contexts in which the {\em Lam\'e equation} arises, 
such as, e.g.,  bond-order and charge density wave systems~\cite{horovitz}, 
nonlinear elasticity~\cite{zhurav},  
phase slips in superconductors and vortex oscillations in a Josephson 
junction~\cite{vortex}, magnetoelastic interaction on curved surfaces 
and symmetric monopoles~\cite{monopole}.  The Lam\'e equation 
appears in systems with a single periodicity, for instance a 
one-dimensional elastic solid comprising identical atoms with a certain 
strength of the potential~\cite{sutherland}, or Hartree equation for 
spinless fermions, or Hartree-Fock equation for spin-1/2 fermions~\cite{shastry}.  
It also arises in field theory contexts such as 
sphaleron solutions of the Abelian Higgs model~\cite{higgs}, in 
connection with elliptic Calogero-Moser systems~\cite{calogero}, and 
in group theoretical aspects of algebraic potentials~\cite{iachello}.

Our goal here is to address the problem of the statistical 
mechanics of thermally controlled excitations in a nonlinear system 
with a Lam\'e potential. The  presented results 
find applications in the various classes of 
physical systems enumerated above.  
Note, however, that  the Lam\'e equation is a {\em quasi-exactly solvable} problem, 
which means that one cannot obtain analytically the complete band spectrum of the 
corresponding Schr\"odinger equation, but only a part of the spectrum, and that  
only for certain values 
of the parameters of the potential, as it will be discussed below.
Therefore, it is not possible to obtain analytically  the corresponding partition function for 
arbitrary temperatures, but only for a set of temperatures. 

The plan of the paper is as follows.  In Sec.~2 we describe the 
soliton-like solutions of the Lam\'e potential.  For the sake of 
completeness, we review some previously obtained 
results~\cite{bena} and then present a new family of kink 
lattice solutions. We also address the zero-temperature ($T=0K$) 
``thermodynamics" of 
the kink crystals and discuss their stability. The single 
topological kink solution is also described, as well as the 
asymptotic interaction energy between kinks. The sine-Gordon results 
are recovered in an appropriate limit.  Section~3 describes an 
approximate method of treating the low-temperature statistical 
mechanics of this kink-bearing system, namely the ideal kink gas 
phenomenology. In Sec.~4 we present the transfer operator formalism 
for an exact calculation of the grand-canonical partition function of 
the system, with some computational details relegated to the Appendix. 
Finally, conclusions and perspectives are presented in Sec.~5.

\section*{2. Soliton Solutions of the Lam\'e Potential}

\begin{figure}[htb]
\begin{center}
\psfrag{f}{{\huge{\bf$\hspace{-0.15cm}{{\phi}}$}}}
\vspace{1cm}
\includegraphics[width=\textwidth]{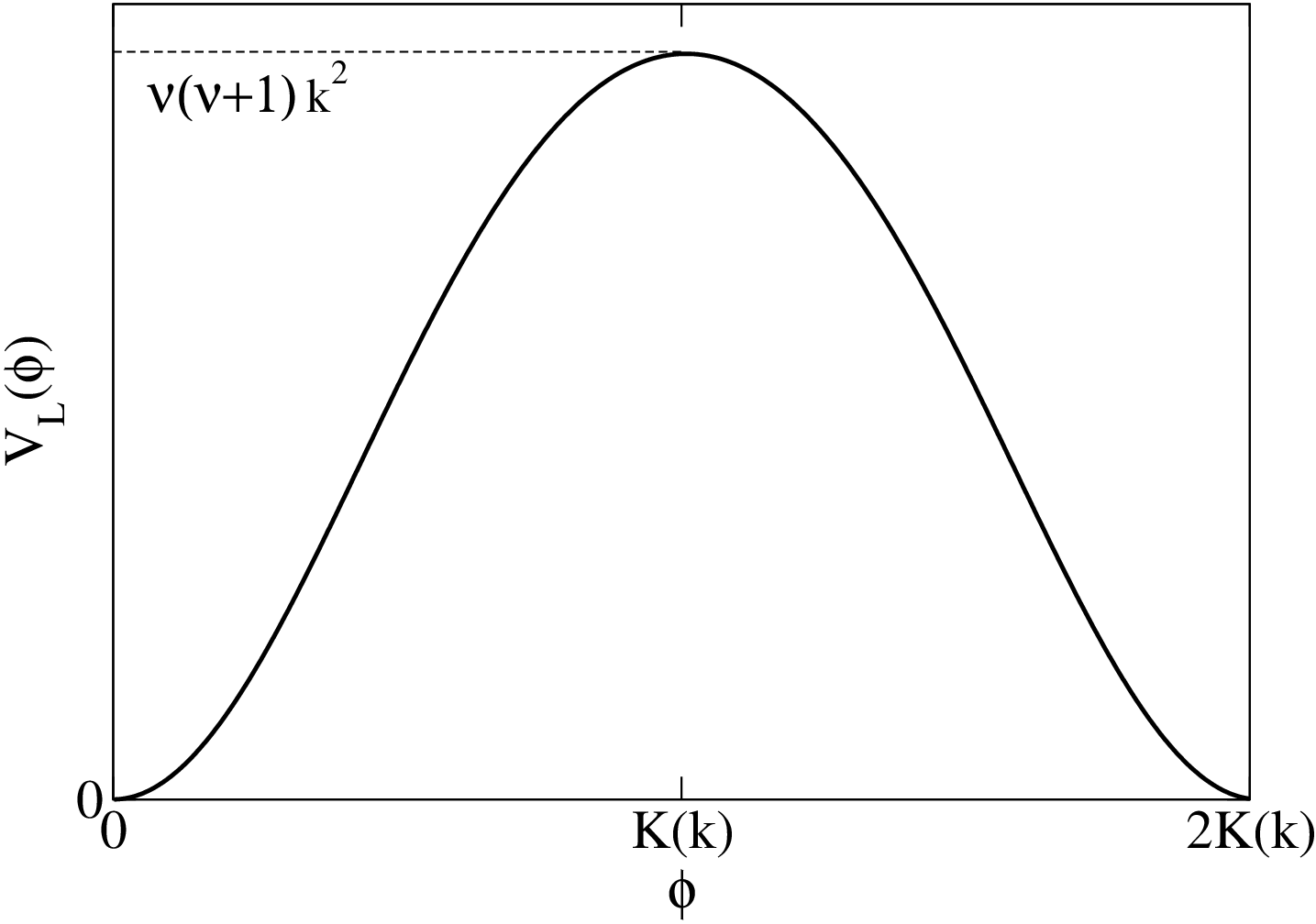}
\end{center}
\caption{}
\vspace{1cm}
\label{figpotential}
\end{figure}

The {\em Lam\'e potential} \cite{magnus,khare,ganguly} 
with a dimensionless parameter $\nu$ is defined as
\begin{equation}
V_{L}(\phi)=\nu (\nu +1)\, k^2\,\mbox{sn}^2(\phi,\,k)\,,
\label{Lame}
\end{equation}
where $\nu(\nu+1)>0$ for the situations of physical relevance.
Here $\mbox{sn}(\phi,\,k)$ is the 
sine-amplitude Jacobi elliptic function of real  modulus 
$k$ ($0\,\leqslant\,k\,\leqslant 1$) and  period $4\,K(k)$; 
$K(k)$ denotes the complete elliptic 
integral of the first kind, see Refs. \cite{gradshteyn,byrd}. 
Lam\'e potential is periodic with a period $2\,K(k)$,
\begin{equation}
V_L(\phi + 2K(k))=V_L(\phi)\,,
\label{period}
\end{equation}
and has one minimum $V_{min}=0$ at $\phi=0$ 
and one maximum $V_{max}=\nu(\nu+1)k^2$ at $\phi=K(k)$, 
see Fig.~\ref{figpotential}.


Consider a classical dimensionless scalar 
field $\phi(x,\,t)$ with $V_L(\phi)$ as the interaction  energy
density. Its dynamics is described by the 
Euler-Lagrange  equation 
\begin{equation}
\displaystyle\frac{\partial^2 \phi}{\partial t^2}\,-\,
c^2\displaystyle\frac{\partial^2\phi}{\partial x^2}\,=\,-\,\omega_0^2
\displaystyle\frac{\partial V_{L}}{\partial \phi}\,
\label{eveq}
\end{equation}
corresponding to a Hamiltonian density
\begin{equation}
\mathcal{H}(x,\,t)=A\left[\frac{1}{2}\left(c^2\phi_x^2+\mathcal{P}^2\right)+\omega_0^2V_L(\phi)\right]\,.
\label{hamden}
\end{equation} 
Here $\mathcal{P}(x,t)=\partial \phi/\partial t \equiv \phi_t$ is the conjugate 
field momentum density, and $\phi_x \equiv \partial \phi/\partial x$.
In view of the periodicity of the Lam\'e potential, 
the field $\phi$ is defined modulo $2K(k)$.
The velocity $c$, the frequency $\omega_0$, and the constant $A$ (that has the dimensions 
[energy]$\cdot$[time]$^2/$[length]) are the parameters of the system defining
the characteristic time $(1/\omega_0)$, length $(c/\omega_0)$, and 
energy $(Ac\omega_0)$ scales.
In the following we shall use dimensionless quantities based on this scaling.

The total energy
\begin{equation}
E_{\Lambda}[\phi,\, \mathcal{P}]=\int_{-\Lambda}^{\Lambda} \mathcal{H}(x,t)\,dx
\label{energy}
\end{equation}
and the topological charge (or the winding number) of the system
\begin{equation}
W_{\Lambda}=\frac{1}{2 K(k)} \int_{-\Lambda}^{\Lambda}\frac{\partial \phi}{\partial x}\,dx=
\frac{1}{2 K(k)}\left[\phi\vline_{\;x=\Lambda}-\phi\vline_{\;x=-\Lambda}\right]
\label{charge}
\end{equation}
are {\em constants of the motion} for fixed boundary conditions ($2\Lambda$ is the length of the system, $-\Lambda \leqslant x \leqslant \Lambda$).
In the ``thermodynamic limit"
$\Lambda \rightarrow \infty$, as seen below, the topological charge is simply the difference between the 
numbers of kinks and  antikinks  in the system.

In the stationary case  the equation for the field
$\phi(x)$ reduces simply to
\begin{equation}
\displaystyle\frac{d^2\phi}{d x^2}\,=\,
\displaystyle\frac{\partial V_{L}}{\partial \phi}\,,
\label{stateq}
\end{equation} that can be integrated
by quadratures, see below.
In view of the covariant form of Eq.~(\ref{eveq}), 
the time-dependent solutions $\phi(x,t)$ are immediately obtained from 
$\phi(x)$ by a ``Lorentz
boosting" to velocity $v$, i.e., $x\rightarrow(1-v^2)^{-1/2}(x-vt)$. 

There are two families of kink lattice 
solutions of (\ref{stateq}). 
From the perspective of the dynamical systems theory, the first family 
of non-topological kinks corresponds to closed periodic orbits in the 
$\left(\phi\,,\phi_x\right)$ phase plane,
while the second family of topological kinks corresponds to open 
periodic orbits in the phase plane.
These different kink lattices reduce to the same single soliton solution in 
the appropriate limit.  

\subsection*{2.1. First family of kink lattice solutions}
 
A first type of stationary solutions is given by
\begin{equation}
\pm\,\sqrt{2}\,(x\,-\,x_0)\,=\,\displaystyle\int_{\displaystyle\phi(x_0)}^
{\displaystyle\phi(x)}
\displaystyle\frac{d{\phi}'}{\sqrt{V_{L}({\phi}',\,k)\,-\,\nu(\nu+1)\,k^2\,a^2}}\,,
\label{quadrature1}
\end{equation}
where $x_0$  and $a^2$ are the two  integration
constants.  We fix $x_0$ through the convention $\phi(0)=K(k)$.
The
control parameter
$a^2$, with $0\leqslant a^2 < 1$,  
determines the appropriate integration domain in Eq. (\ref{quadrature1}),
the profile and the properties 
of the field $\phi(x)$.  

\begin{figure}[htb]
\begin{center}
\psfrag{(a)}{\Large\bf\hspace{-0.2cm}(a,k)}
\psfrag{f}{{\huge{\bf$\hspace{-0.3cm}{{\phi}}$}}}
\vspace{1cm}
\includegraphics[width=\textwidth]{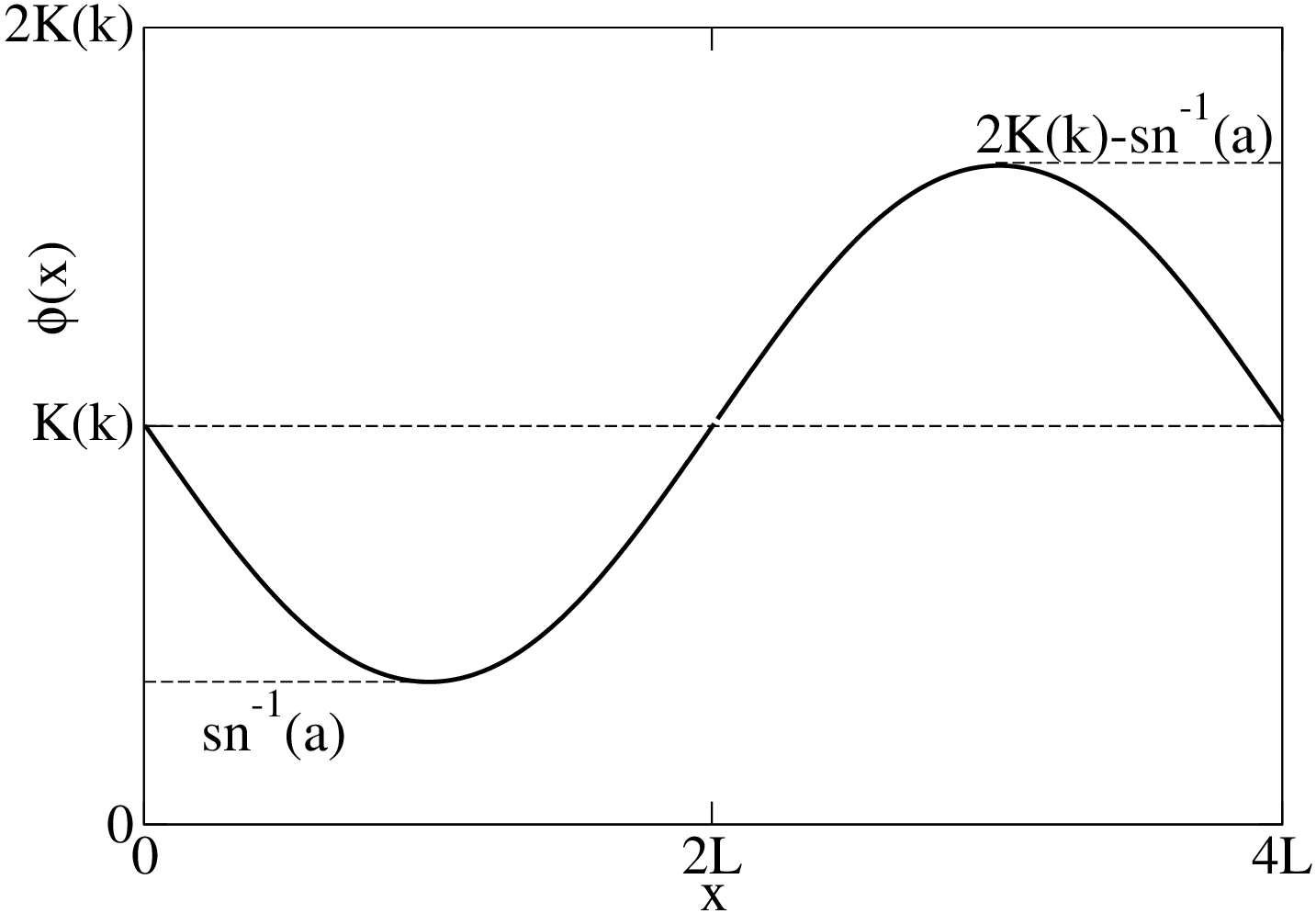}
\end{center}
\caption{}
\vspace{1cm}
\label{figlattice1}
\end{figure}

When extended to the whole real axis $\Lambda \rightarrow \infty$, 
Eq.~(\ref{quadrature1}) leads to a
{\em soliton lattice} solution:
\begin{equation}
\mbox{sn}^2(\phi,\,k)=\frac{(1-k^2a^2)\,-\,(1-a^2)\,\mbox{sn}^2(y,\,t)}
{(1-k^2a^2)-k^2(1-a^2)\,\mbox{sn}^2(y,\,t)}\,,
\label{solution1}
\end{equation}
where $t\equiv \sqrt{(1-a^2)/(1-k^2a^2)}$ and 
$y\equiv \sqrt{2\nu(\nu+1)k^2(1-k^2a^2)}\,x$. 
This represents an array of {\em nontopological} solitons, i.e., a periodic sequence 
of ``small" (nontopological) kink--antikink pairs.
In Fig.~\ref{figlattice1} we show a typical profile of this type of soliton lattice solution 
(i.e., the field $\phi$ as a function of $x$) for a fixed value of $a^2$. 
The spatial size of one kink/antikink (which represents half of the spatial period of the lattice)
is a function of the control parameter $a^2$
\begin{equation}  
2L\,(a^2)=\frac{2K(t)}{\sqrt{2\nu(\nu+1)k^2(1-k^2a^2)}}\,,
\label{sizekink1}
\end{equation}
and the topological charge of such a ``small" kink/antikink is, respectively, 
$\pm \left[1- \mbox{sn}^{-1}(a,k)/K(k)\right]$.


The energy per kink/antikink of the lattice, as a function of $a^2$, 
can be computed as
\begin{equation}
E_{KL}(a^2)=2\,\sqrt{\displaystyle\frac{2\,\nu(\nu+1)k^2}{1\,-\,k^2a^2}}\,
\left[\left(\frac{1}{k^2}-\frac{a^2}{2}\right)\,K(t)\,-\,\frac{k'^2}{k^2}\,
\Pi(k^2 t^2,\,t)\right]\,,
\label{esl1}
\end{equation}
where  $\Pi(k^2 t^2,\,t)$ denotes the complete elliptic integral 
of the third kind \cite{gradshteyn,byrd}.

\subsubsection*{2.1.1. Kink lattice  ``thermodynamics" at $T=0K$}

One can immediately compute the thermodynamic characteristics of
the ground state of this classical kink-antikink crystal. 
The ground-state energy per kink is expressed by Eq.~(\ref{esl1}), 
while the specific volume per kink is given by Eq.~(\ref{sizekink1}).
Correspondingly, the thermodynamic pressure  
is obtained as
\begin{equation}
P_{KL}=-\left(\frac{\partial E_{KL}}
{\partial (2L)}\right)_{T=0K}=-\frac{dE_{KL}/d(a^2)}{d(2L)/d(a^2)}=-\nu(\nu+1)k^2a^2\,.
\label{pkl1}
\end{equation} 
(in deriving the last equality we have used the properties of the 
derivatives of the elliptic 
functions~\cite{gradshteyn,byrd}).
Thus, the internal energy of the crystal increases with increasing volume
under these isothermal conditions.
The equation of state can  be  obtained formally
by eliminating the parameter $a^2$ between 
Eqs.~(\ref{sizekink1}) and (\ref{pkl1}). 

\begin{figure}[htb]
\begin{center}
\vspace{1cm}
\includegraphics[width=\textwidth]{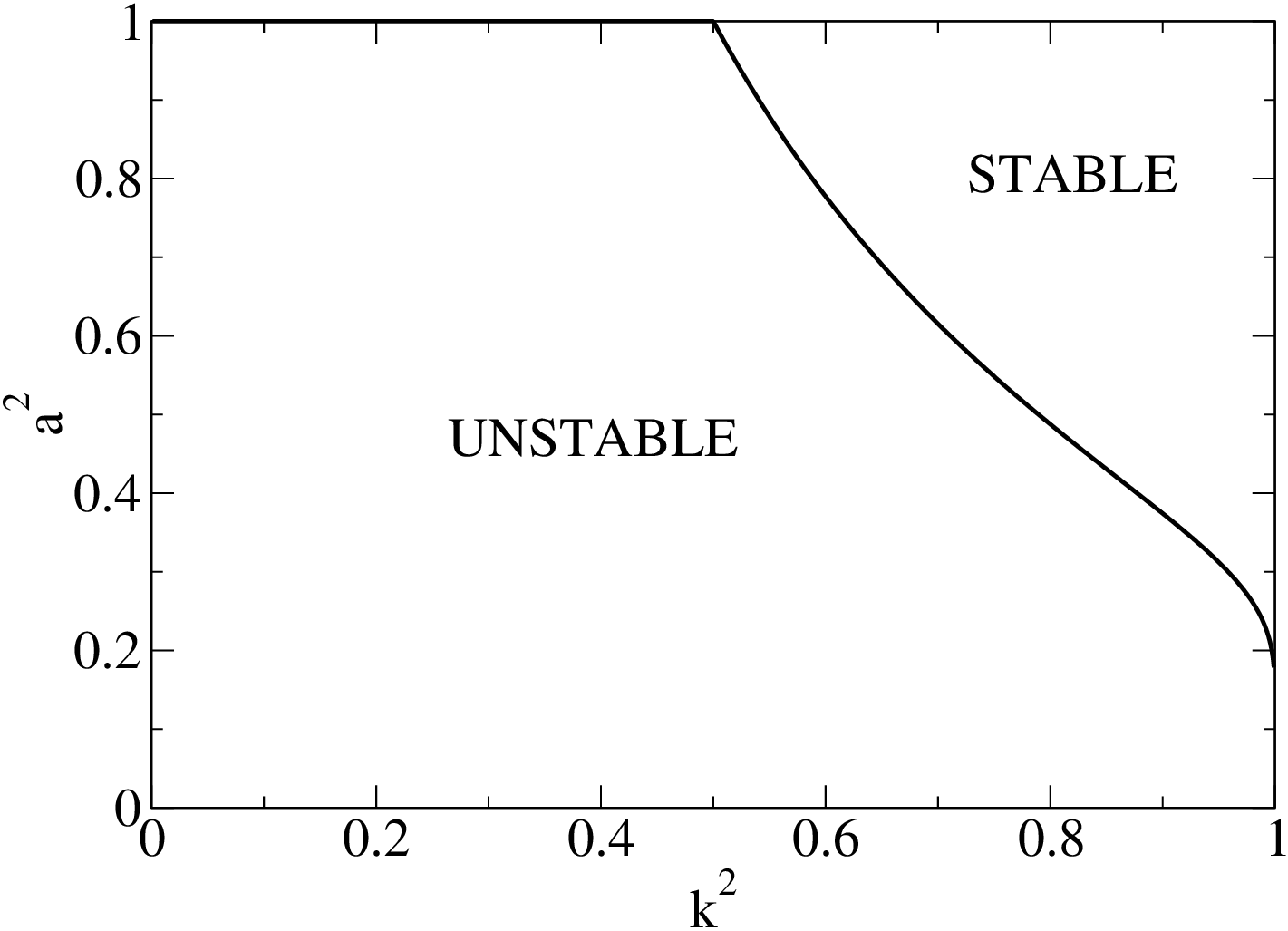}
\end{center}
\caption{}
\vspace{1cm}
\label{figstability}
\end{figure}

The chemical potential at zero temperature $\mu_{KL}$ 
is just the enthalpy per kink,
\begin{eqnarray}
&&\mu_{KL}=E_{KL}+2L P_{KL}=2 \sqrt{\frac{2 \nu(\nu+1)k^2}{1-k^2a^2}}
\left[\frac{1-k^2a^2}{k^2}K(t)-\frac{{k'}^2}{k^2}\Pi(k^2t^2,t)\right]\, ,
\nonumber\\
&&
\end{eqnarray}
where the complementary modulus $k'^2=1-k^2$. 
Finally, the isothermal compressibility
at $T=0K$ is
\begin{equation}
\chi_{KL}=-\frac{1}{2L}\frac{\partial (2L)}{\partial P_{KL}}=-\frac{1}{2L}\frac{\displaystyle {d(2L)}/{d(a^2)}}
{\displaystyle{dP_{KL}}/{d(a^2)}}=\frac{K(t)-(1/a^2)\, E(t)}{2\nu(\nu+1)k^2(1-a^2)K(t)}\,.
\end{equation}
Here $K(t)$ and $E(t)$ are, respectively, the first- and second-kind
complete elliptic integrals of modulus $t$, see~\cite{gradshteyn,byrd}.
One notices that $\chi_{KL}$ can change sign upon variation of the parameters
$k^2$ and $a^2$, the states with $\chi_{KL}<0$ being ``mechanically" unstable.
The corresponding stability
diagram of this kink lattice solution at $T=0K$ 
in the plane of the parameters 
$(k^2,\,a^2)$ is presented in Fig.~\ref{figstability}. 


\subsubsection*{2.1.2. The sine-Gordon limit}

As discussed in~\cite{bena}, the Lam\'e potential~(\ref{Lame})
reduces to the sine-Gordon (sG) potential in the limit 
\begin{equation}
k \rightarrow 0\,,\quad
\nu(\nu+1)\rightarrow \infty\,,\quad \mbox{with}\quad 
\nu(\nu+1) k^2\rightarrow Q=\mbox{finite}\,.
\label{sglimit}
\end{equation}
In this limit the above nontopological kink lattice solution reduces to
$\phi_{sG}(x)$:
\begin{equation}
\mbox{sin}^2(\phi_{sG})=1-(1-a^2)\mbox{sn}^2(\sqrt{2Q}x,\,\sqrt{1-a^2})=\mbox{dn}^2(\sqrt{2Q}x,\,\sqrt{1-a^2})\,,
\end{equation}
where $\mbox{dn}(\sqrt{2Q}x,\,\sqrt{1-a^2})$ is the delta-amplitude Jacobi elliptic 
function~\cite{gradshteyn,byrd}.
The kink/antikink width is $2L_{sG}(a^2)=2K(\sqrt{1-a^2})/\sqrt{2Q}$, 
and the energy per kink/antikink
\begin{equation}
E_{sGKL}=2\sqrt{2Q}[E(\sqrt{1-a^2})-(a^2/2)K(\sqrt{1-a^2})]\,. 
\end{equation}

The pressure of the sine-Gordon kink lattice at $T=0K$ is $P_{sGKL}=-Qa^2$,
and the chemical potential 
$\mu_{sGKL}=2\sqrt{2Q}[E(\sqrt{1-a^2})-(a^2/2)K(\sqrt{1-a^2})]$.
The isothermal compressibility
\begin{equation}
\chi_{sGKL}=\displaystyle\frac{K(\sqrt{1-a^2})-1/a^2E(\sqrt{1-a^2})}
{2Q(1-a^2)K(\sqrt{1-a^2})}\,,
\end{equation}
is always negative, which means that this class of sG  
kink lattice solution is {\em unstable} (see the $k\rightarrow0$ 
limit in Fig.~\ref{figstability} and~\cite{radha}).

\subsection*{2.2. Second family of kink lattice solutions}

\begin{figure}[htb]
\begin{center}
\psfrag{f}{{\huge{\bf$\hspace{-0.3cm}{\tilde{\phi}}$}}}
\vspace{1cm}
\includegraphics[width=\textwidth]{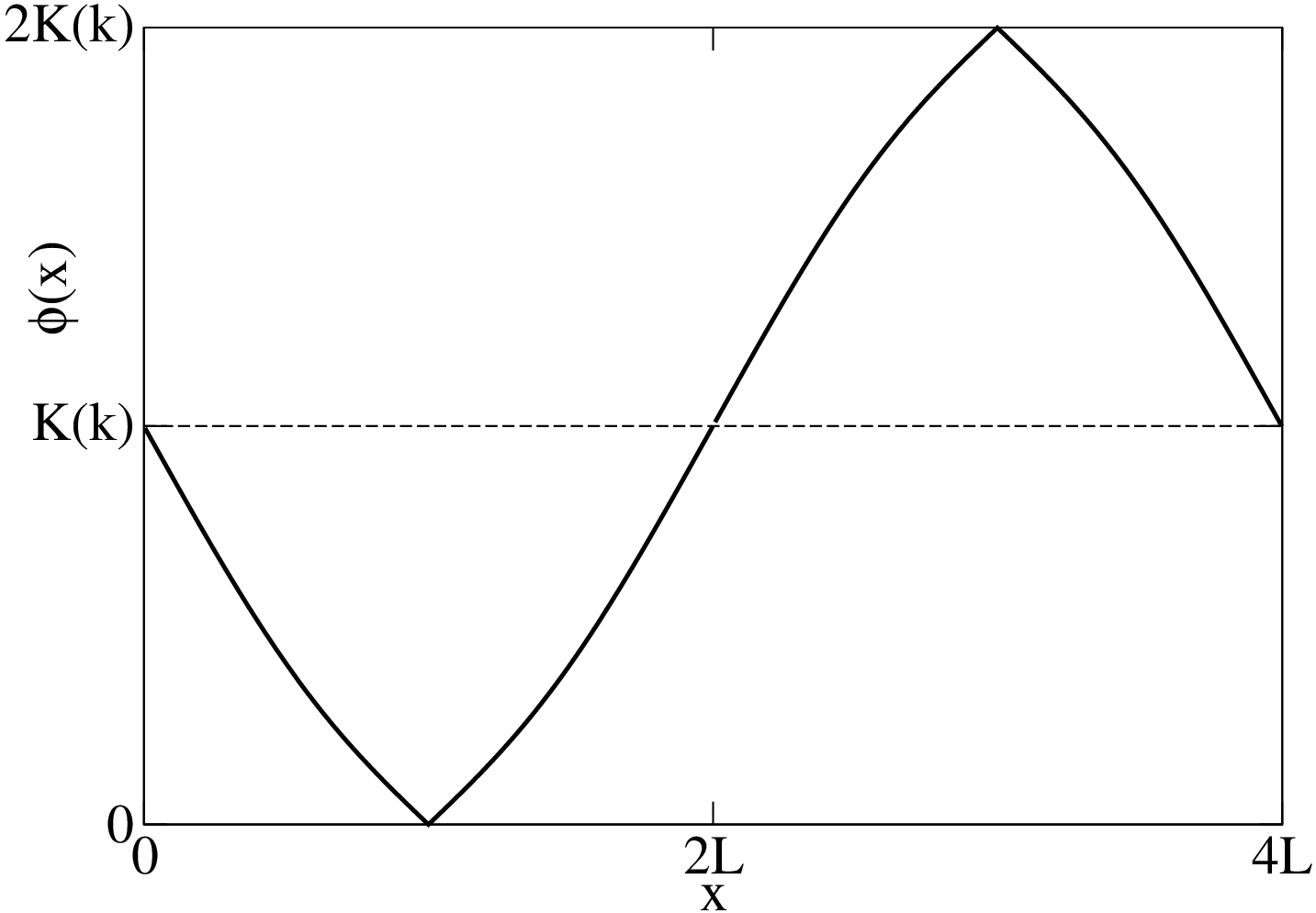}
\end{center}
\caption{}
\vspace{1cm}
\label{figlattice2}
\end{figure}

The second type of stationary solutions is given by
\begin{equation}
\pm\,\sqrt{2}\,(x\,-\,x_0)\,=\,\displaystyle\int_{\displaystyle\tilde\phi(x_0)}^
{\displaystyle\tilde\phi(x)}
\displaystyle\frac{d{\phi}'}{\sqrt{V_{L}({\phi}',\,k)\,+\,\nu(\nu+1)\,k^2\,a^2}}\,,
\label{quadrature2}
\end{equation}
where $x_0$  and $a^2\geqslant 0$ are the two integration
constants.  Here again we fix $x_0$ through the convention $\tilde\phi(0)=K(k)$.
One notices that this family of solutions can be obtained 
{\em formally}
from the previous one, Eq.~(\ref{solution1}), by replacing $a^2$ by $-a^2$, and
thus:
\begin{equation}
\mbox{sn}^2(\tilde{\phi},\,k)=
\displaystyle\frac{1-\mbox{sn}^2(\tilde{y},\,\tilde{t})}
{1-k^2\mbox{sn}^2(\tilde{y},\,\tilde{t})}\,,
\label{solution2}
\end{equation}
where $\tilde{t}=\sqrt{(1+k^2a^2)/(1+a^2)}$ and
$\tilde{y}=\sqrt{2\nu(\nu+1)k^2(1+a^2)}\,x$ with $a^2\geqslant 0$.
This is a {\em topological}~ kink lattice solution, see Fig.~\ref{figlattice2},
i.e., a succession of topological kinks (or antikinks),
of  kink/antikink width (half lattice period)
\begin{equation}
2\tilde{L}(a^2)=\displaystyle\frac{2K(\tilde{t})}
{\sqrt{2\nu(\nu+1)k^2(1+a^2)}}\,.
\end{equation}


The energy per kink/antikink of the lattice is  found to be:
\begin{equation}
\tilde{E}_{KL}=2\sqrt{\displaystyle\frac{2\nu(\nu+1)k^2}{1+a^2}}
\left[\left(\frac{1}{k^2}+\frac{a^2}{2}\right)\,
K(\tilde{t})-\frac{k'^2}{k^2}\,\Pi(k^2,\,\tilde{t})\right]\,.
\end{equation}

\subsubsection*{2.2.1. $T=0K$ ``thermodynamics"}

Proceeding as in the previous case of the first family of kink lattice
solutions, one obtains for the pressure of the ground state 
of the second type lattice solutions
\begin{equation}
\tilde{P}_{KL}=\nu(\nu+1)k^2a^2\,,
\end{equation}
and for the corresponding chemical potential 
\begin{equation}
\tilde{\mu}_{KL}=2\sqrt{\displaystyle\frac{2\nu(\nu+1)k^2}{1+a^2}}
\left[\left(\frac{1}{k^2}+{a^2}\right)\,
K(\tilde{t})-\frac{k'^2}{k^2}\,\Pi(k^2,\,\tilde{t})\right]\, .  
\end{equation}

Finally, the isothermal compressibility
\begin{equation}
\tilde{\chi}_{KL}=\displaystyle
\frac{k^2K(\tilde{t})+(1/a^2)E(\tilde{t})}
{2\nu(\nu+1)k^2(1+k^2a^2)K(\tilde{t})}\,,
\end{equation}
which is always positive, and thus the kink lattice solution of the second type
is stable at $T=0K$.

\subsubsection*{2.2.2. The sine-Gordon limit}

In the sine-Gordon limit~(\ref{sglimit}),
the above kink lattice solution reduces to
\begin{equation}
\mbox{sin}^2(\tilde{\phi}_{sG})=1-\mbox{sn}^2(\sqrt{2Q(1+a^2)}x,\,1/\sqrt{1+a^2})=\mbox{cn}^2(\sqrt{2Q(1+a^2)}x,\,1/\sqrt{1+a^2})\,,
\end{equation}
of spatial period $2\tilde{L}_{sG}(a^2)=2K(1/\sqrt{1+a^2})/\sqrt{2Q(1+a^2)}$,
where $\mbox{cn}(\sqrt{2Q(1+a^2)}x,\,1/\sqrt{1+a^2})$ is the cosine-amplitude Jacobi elliptic 
function~\cite{gradshteyn,byrd}.
The corresponding energy per kink is
\begin{equation}
\tilde{E}_{sGKL}=2\sqrt{\displaystyle\frac{{2Q}}{1+a^2}}\left[
(1+a^2)E(1/\sqrt{1+a^2})-(a^2/2)K(1/\sqrt{1+a^2})\right]\,,
\end{equation}
the chemical potential is
$\tilde{\mu}_{sGKL}=2\sqrt{2Q(1+a^2)}E(1/\sqrt{1+a^2})$, and the
pressure is $\tilde{P}_{sGKL}=Qa^2$. The
positiveness of the isothermal compressibility
$\chi_{sGKL}=E(1/\sqrt{1+a^2})/[2Qa^2K(1/\sqrt{1+a^2})]$ confirms the 
{\em thermodynamic stability} 
of the sG kink lattice of the second type. These results are in agreement
with those found previously in~\cite{gupta} for the sG potential.

\subsection*{2.3. The single soliton solution}

\begin{figure}[htb]
\begin{center}
\psfrag{f}{{\huge{\bf$\hspace{-0.3cm}{{\phi}}$}}}
\vspace{1cm}
\includegraphics[width=\textwidth]{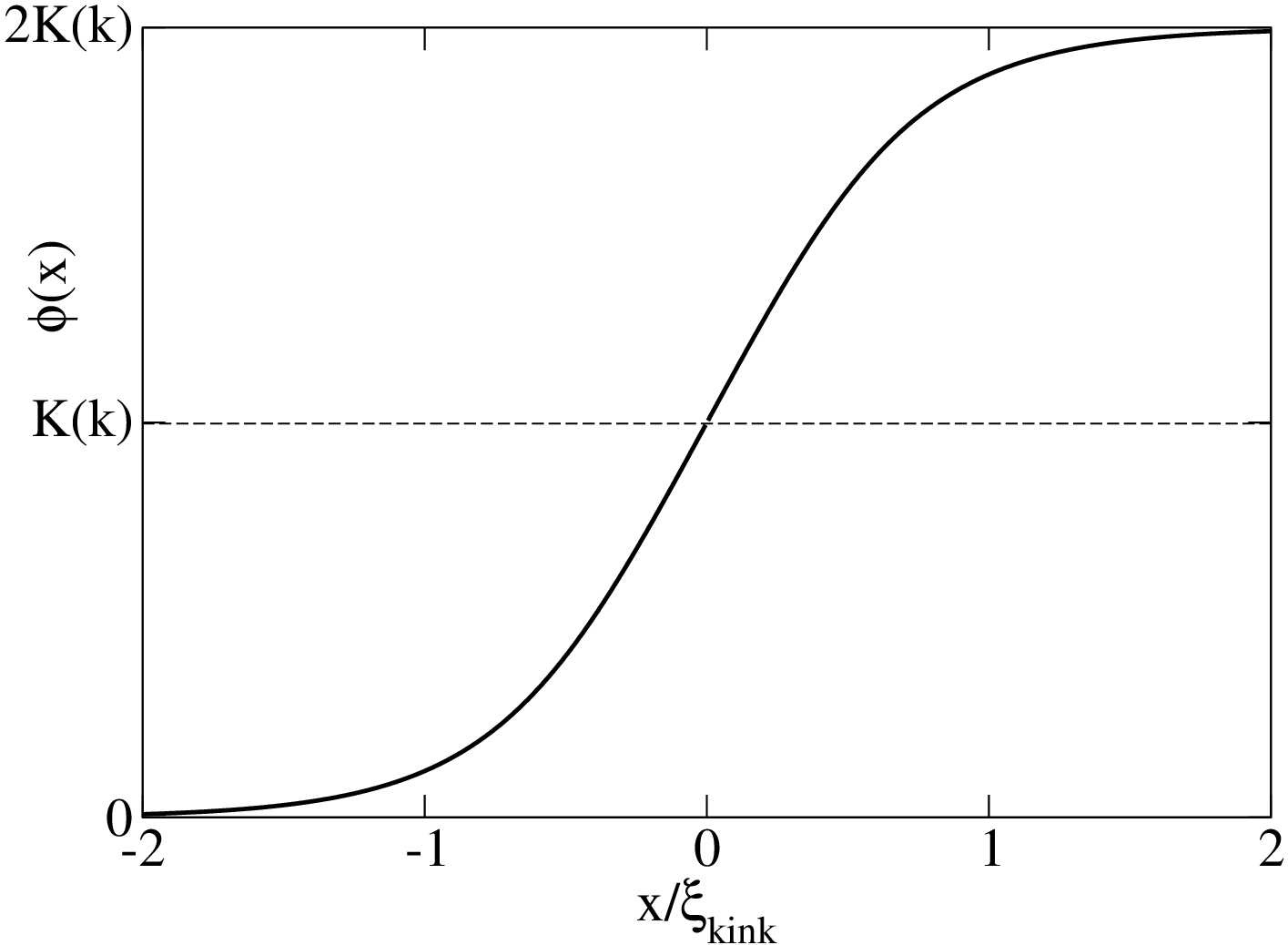}
\end{center}
\caption{}
\label{figkinkprofile}
\end{figure}

For the sake of completeness we consider the single soliton solution 
as well.  The corresponding single soliton solution can be obtained 
from {\em either} of the above soliton lattices in the limit of 
``infinite kink dilution" , i.e., when the period of the lattice  
becomes infinite, $L,\,\tilde{L}\rightarrow \infty$,
which corresponds to the control parameter $a^2 \rightarrow 0$. 
It is a {\em topological}, ``large" kink/antikink of charge 
\begin{equation}
\frac{1}{2K(k)}\int_{-\infty}^{\infty}\phi_x\,dx=\pm1\,,
\end{equation}
given by
\begin{equation}
\mbox{sn}(\phi,\,k) =\pm \frac{1}{\sqrt{1+k'^2\,\mbox{sinh}^2(y^*)}}\, ,
\label{kink}
\end{equation}
where $y^{*}=\sqrt{2\,\nu(\nu+1)\,k^2}\,x$ and $k'^2=1-k^2$. The characteristic
``spatial extent"
of such a soliton is  $\xi_{kink}=2/\sqrt{2\nu(\nu+1)k^2}$.
In Fig.~\ref{figkinkprofile} we present the typical profile of a single kink.


The energy (rest mass) of the static single kink/antikink 
is shown \cite{bena} to be  
\begin{equation}
E_K=\sqrt{2\,\nu(\nu+1)}\;\mbox{ln} \left(\frac{1+k}{1-k}\right)\,.
\end{equation}
In view of the covariant form of equation~(\ref{eveq}),
a kink moving with a velocity $v$ has an energy
\begin{equation}
E_K(v)=(1-v^2)^{-1/2}E_K=(E_K^2+\mathcal{P}^2)^{1/2}\,, 
\label{ekv}
\end{equation}
where $\mathcal{P}=E_K\,v(1-v^2)^{-1/2}$ is the ``relativistic" momentum.

One can also compute the energy of the interaction of one kink/antikink 
in the lattice with the rest of the lattice as $U(a^2)=E_{KL}(a^2)-E_K$ 
(for the first type kink lattice) or as  
$\tilde{U}(a^2)=\tilde{E}_{KL}(a^2)-E_K$ (for the second type kink lattice).
From these expressions one can deduce 
(through a Taylor expansion in the parameter $a^2$) 
the {\it long-distance}  interaction energy between two kinks.

Note that there is another, more general way to obtain the asymptotic 
interaction energy  between two solitons -- Manton's method 
\cite{bena,manton}. Given the asymptotic shape of the single 
kink~(\ref{kink}), 
that in our case reads (e.g., for $x \rightarrow \infty$) 
\begin{equation}
\phi_{as} \approx \frac{2}{\sqrt{1-k^2}} \;
\mbox{exp}\left(-\sqrt{2 \nu (\nu+1)k^2} x \right)\,,
\end{equation}
one obtains for the asymptotic interaction energy between two solitons 
separated by a distance $D \gg 1/\sqrt{2 \nu (\nu+1)k^2}$:
\begin{equation}
U_{as}(D) \approx \pm {\frac{8 \sqrt{2 \nu (\nu+1)k^2}}{1-k^2}} \; 
\mbox{exp} \left(-\sqrt{2 \nu (\nu+1)k^2} D \right)\,.
\end{equation}
The positive sign corresponds to a  repulsive interaction between solitons 
of the same topological charge, while between a kink and an antikink the 
interaction is attractive. For the particular case of the 
first type of soliton lattice, in the limit of ``high kink dilution" 
($a^2 \ll 1$), one has 
\begin{equation}
D=2L (a^2)\approx 1/\sqrt{2 \nu 
(\nu+1) k^2}\; \mbox{ln}\left[\frac{16}{a^2 (1-k^2)}\right]\;\gg\;1/\sqrt{2 \nu (\nu+1)k^2}\,,
\end{equation}
and thus the asymptotic attractive interaction energy between a kink and 
an antikink in the lattice is
\begin{equation}
U_{as}(a^2)\approx - a^2\;\sqrt{\nu(\nu+1)k^2/2}\,.
\end{equation}
The same expression is obtained in the limit of ``high kink dilution" 
for the second type of soliton lattice.  Finally, there is no difficulty 
or surprise in obtaining the properties of the sG single soliton in 
the appropriate limit~(\ref{sglimit}), see also~\cite{gupta}.

Next, let us briefly consider the question of the behavior of the single 
kink solution of the Lam\'e potential with respect to small perturbations 
of its shape. For small oscillations $\psi(x)\,\exp(i \omega t)$ around 
the stationary kink solution~(\ref{kink}), the linearized dynamics 
equation reads
\begin{equation}
\left\{-\frac{d^2}{dx^2} + 2 \nu(\nu+1)k^2\left[1-2(1+k^2)
\mbox{sn}^2(\phi,k)+3k^2\mbox{sn}^4(\phi,k)\right]-\omega^2\right\}\,\psi(x)=0\,.
\label{stabilitykink}
\end{equation}
Consider the following change of variables:
\begin{equation}
z=\frac{2}{(1-k^2) \mbox{cosh}(2 y^*) +(1+k^2)}\,,
\end{equation} 
and
\begin{equation}
\psi=\displaystyle\left[({1-k^2})\,z/2\right]^s \,\chi(z)\,,\quad
\quad
s \equiv \sqrt{\frac{1}{4}-\frac{\omega^2}{8 \nu(\nu+1)k^2}}\,.
\end{equation}
With these new variables, the stability equation~(\ref{stabilitykink}) 
becomes Heun's 
equation~\cite{brihaye}:
\begin{eqnarray}
\frac{d^2\chi}{dz^2}&+&\left(\frac{2s+1}{z}+\frac{1/2}{z-1}+
\frac{1/2}{z-1/k^2}\right)\frac{d\chi}{dz}+\nonumber\\
&+&\frac{(s-1/2)(s+3/2)z-(s+1)(2s-1)(1+k^2)/(2k^2)}{z(z-1)(z-1/k^2)}\,\chi=0\,,
\end{eqnarray}
that is a subject of current intensive research in mathematical 
physics~\cite{maier2}.
Further on, following~\cite{brihaye}, one can make a new change of variables 
\begin{equation}
z=\mbox{sn}^2(u,\,k)\,, \qquad f(u)=[\mbox{sn}(u,\,k)]^{(4s+1)/2}\,\chi(z(u))\,,
\end{equation}
and the Heun equation above reduces to the Schr\"odinger equation
\begin{equation}
-\displaystyle\frac{d^2f(u)}{du^2}+\tilde{V}(u)f(u)=hf(u)
\end{equation}
with the ``potential" $\tilde{V}(u)=(15/4)k^2\,\mbox{sn}^2(u,\,k)+(16s^2-1)/[4\;\mbox{sn}^2(u,\,k)]$
and a fixed ``energy" $h=9(1+k^2)/4$.

As explained in detail in, e.g.,~\cite{currie}, 
and also checked directly, these stability
equations always admit the Goldstone $\omega=0$ ($s=1/2$) mode as a solution,
thus restoring the translational invariance of the original
Hamiltonian~(\ref{hamden}) broken by the introduction of the kink into the
system. In addition to the translation mode at $\omega=0$, although difficult
to find exactly, there may exist
additional {\em bound states} with nonzero frequency of the stability equations.
These solutions represent {\em internal oscillations of the kink} (or shape 
modes), corresponding to harmonically varying kink shape around the kink 
center. They are strongly localized around the static kink.
Besides these discrete bound state solutions, there exist continuum states
(extended modes) corresponding to phonons (see below Sec.~3.2) 
that are scattered by the kink (asymptotically, i.e., far from kink's center, 
this scattering results simply in a phase shift as compared to the kink-free 
system). Note that the degrees of freedom of the bound states (internal modes 
of the kink) are obtained at the expense of the phonon field (whose local 
density is altered by the presence of the kink), i.e., the excitation of 
these internal modes can be seen as a `capture' of phonons by the kink, 
see~\cite{currie} for a detailed discussion of this point.

One could legitimately address the question of the linear stability  
in the case of the kink lattice solutions, too. However, the corresponding
linearized evolution equation of the perturbation (although reminiscent of 
a Heun's equation) is more complicated and will not be discussed further.

\section*{3. Finite-temperature thermodynamics: The ideal kink  gas approximation}

\subsection*{3.1. The contribution of the kinks}

A first approach to the study of the finite-temperature $T>0$ $K$ 
thermodynamics of the field $\phi$ can be discussed within the framework 
of the {\em ideal kink gas approximation},
see~\cite{currie,deleonardis,krumhansl} for a more detailed description of the method. 
This approximation consists of assuming that, {\em at sufficiently low temperatures}, 
very few kink-like excitations are present, and thus
the mean distance between them is very large as compared to their typical size $\xi_{kink}$. 
Therefore, these kinks and antikinks behave like {\em independent, non-interacting particles},
with an energy (when moving with velocity $v$) given by Eq.~(\ref{ekv}). 
Moreover, the interaction between individual kinks
and the phonons (see below) is also neglected, i.e., kinks and 
phonons contribute independently to the thermodynamic properties of the system.

Let us concentrate first on the contribution of the kink gas.
Suppose there are $N_K$ kinks and $N_{\overline{K}}$ antikinks in the system; 
then $W_{\Lambda}=N_K-N_{\overline{K}}$ 
is the constant topological charge (winding number), Eq.~(\ref{charge}). 
There is no restriction on the ordering of 
these kinks and antikinks along the system. 
Let $N_{\Lambda}=N_K+N_{\overline{K}}=2N_K-W_{\Lambda}$ be the 
total number of kink-like excitations in the system. 
Then the {\em grand-canonical} partition function is simply:
\begin{equation}
\Xi_{\Lambda}(\beta,\mu)=\sum_{N_{\Lambda}=0}^{\infty}\exp(\mu\beta N_{\Lambda})Z_{\Lambda}(\beta)\,,
\end{equation}
where $\mu$ is the chemical potential associated with the creation of a kink-like excitation in the system, $T$ is the temperature,
$\beta=1/(k_BT)$, and $Z_{\Lambda}(\beta)$
is the canonical partition function,
\begin{equation}
Z_{\Lambda}(\beta)=\displaystyle\frac{(2\Lambda)^N}
{N!h^N}\left\{\int_{-\infty}^{+\infty}d\mathcal{P}\exp
\left[-\beta(E_K^2+\mathcal{P}^2)^{1/2}\right]\right\}^N
\end{equation}
(with $h$  Planck's constant).
Thus, finally  
\begin{eqnarray}
\Xi_{\Lambda}(\beta,\mu)&&=\exp\left[(4E_K \Lambda/h)K_1(\beta E_K)
\exp(\beta \mu)\right]\nonumber\\
&&\approx 
\exp\left\{(2\Lambda/h)(2\pi E_K/\beta)^{1/2}\exp[\beta (\mu-E_K)]\right\}
\end{eqnarray}
with the approximation holding for low temperatures $\beta \gg 1$
($K_1$ is the modified Bessel function~\cite{gradshteyn}). This leads to a pressure 
\begin{equation}
P_K=(2\pi E_K/\beta^3 h^2)^{1/2}\exp[\beta (\mu-E_K)] , 
\end{equation}
and to an average kink-density (i.e., number of kinks per unit length)
\begin{equation}
n_K=(\partial P_K/\partial \mu)_T=(2\pi E_K/\beta h^2)^{1/2}\exp[\beta (\mu-E_K)]. 
\end{equation}
The situation of physical relevance is that when there is no external constraint
on the number of kink-like excitations in the system, i.e., 
the kink density is solely determined by the temperature;
this amounts to setting  $\mu=0$ in the above equation, thus obtaining
\begin{equation}
n_K=(2\pi E_K/\beta h^2)^{1/2}\exp(-\beta E_K)\,,
\label{kden}
\end{equation}
which shows that the kink density  is {\em exponentially small} 
at low temperatures $\beta \gg 1$.

The free energy density (i.e., the free energy per unit length
of the system) of this ideal kink gas is then simply 
\begin{equation}
F_{K}=-(1/\beta)\, n_K\,,
\end{equation}
and the internal energy density
\begin{equation}
U_{K}=\partial(\beta F_{K})/\partial \beta=\left[E_K-1/(2\beta)\right]\,n_K\,.
\end{equation}
Finally, the specific heat per unit length is
\begin{equation}
C_{K}=k_B\left[(\beta E_K-1/2)^2-1/2\right]\,n_K\,.
\end{equation}

In analogy with the correlation functions for periodic 
potentials~\cite{deleonardis,scalapino}, e.g. $\langle\sin\phi(0), 
\sin\phi(x)\rangle$ for the sine-Gordon case, an appropriate measure 
of the static field-field spatial correlation function is given by the 
following expression:
\begin{equation}
\mathcal{C}(x)=\langle \mbox{sn}(\phi(0),k)\,\mbox{sn}(\phi(x),k)\rangle\,,
\label{calx}
\end{equation}
where $\langle ...\rangle$
designates the usual thermal average (in agreement with the convention in Sec.~II,
$\phi(0)=K(k)$).
The choice of the function $\mbox{sn}(\phi, k)$ for the study 
of the spatial correlations of the field $\phi$ is related to its
 ``sensitivity" to the presence of a kink
(antikink); indeed, if a kink (i.e., a ``jump'" of $2K(k)$ in the field $\phi$) is present on the segment 
of length $x$, this leads to a change in the sign of $\mbox{sn}(\phi,k)$. 
The value of $\mathcal{C}(x)$
can be easily estimated using the following qualitative 
argument~\cite{deleonardis}: at low kink density 
the probability $\mbox{Prob}(n,x)$ of finding $n$ kinks in a segment of length $x$ is simply a Poisson distribution
of mean $\langle n \rangle = n_K x$, i.e.,
\begin{equation}
\mbox{Prob}(n,x)= \frac{(n_K x)^n}{n !}\,\exp(-n_K x)\,. 
\end{equation}
Correspondingly, $\mbox{sn}(\phi(0),k)\,\mbox{sn}(\phi(x),k)=(-1)^n {sn}(\phi(0),k)=(-1)^n$. Thus the correlation function is simply
\begin{equation}
\mathcal{C}(x)=\sum_{n=0}^{\infty}(-1)^n\,\mbox{Prob}(n,\,x)=\exp(-2n_K\,x)\,,
\end{equation}
which corresponds to a correlation length $\xi_K=1/(2n_K)$ of the field $\phi$.

\subsection*{3.2. Linear modes (phonons)}
\label{phonons}

Let us briefly turn to the description of the phonons, i.e., the
small-amplitude harmonic vibrations in the Lam\'e potential well.
The equation of motion (\ref{stateq}) for the field $\phi$ 
can be linearized around the minimum $\phi=0$ of the potential 
$V_L(\phi)$, thus leading to the following dispersion relation for the phonons:
\begin{equation}
\omega_q^2=q^2+2\nu(\nu+1)k^2\,.
\label{dispersion}
\end{equation}
All the frequencies are real for all $q \geqslant 0$.
In the semiclassical approximation,
the contribution of the phonons to the free energy {\em per unit length}
of the system at a temperature $T$ is thus given by
\begin{equation}
F_{ph}=\frac{1}{\beta \,\Delta x} \mbox{ln} \left(\frac{h  \beta}
{2 e\,\Delta x}\right) + \frac{1}{2 \beta}\,\sqrt{2 \nu(\nu+1)k^2}\,,
\label{fvib}
\end{equation}
where $\Delta x$ is the size of a spatial cell corresponding to the 
smallest-wavelength linear mode of the system, i.e., to the Debye 
cut-off in terms of frequency
(usually, $\Delta x$ corresponds to the unit-cell length of the 
corresponding discrete lattice of the system under study). 
Taking the continuum limit $\Delta x \rightarrow 0$ leads to the 
well-known ``ultraviolet catastrophe". 
The contribution of the phonons to the internal energy density is then
\begin{equation}
U_{ph}=\partial(\beta \,F_{ph})/\partial \beta=1/(\beta\, \Delta x)\,,
\end{equation}
while the specific heat per unit length is a constant, $C_{ph}=1/\Delta x$.
No interaction of the phonons with the kinks is taken into account
in the above thermodynamic expressions.

The total pressure, free energy density, internal energy density, 
and specific heat per unit length of the classical field $\phi$
is the sum of the independent contributions of the phonons and the
ideal kink gas, respectively. 
In a more refined approach~\cite{currie}, one can take into account
the interaction between phonons and individual kinks as a `trapping' of phonons
by the kinks, i.e., the excitation of the internal bound modes of the kinks at 
the expense of the
degrees of freedom of the phonon field.
However, this results only in a transfer of energy between the linear modes 
(phonons) and nonlinear ones (kinks), i.e., it does not affect the global 
thermodynamic properties of the system inside the limits of the 
approximate ideal kink gas scheme.

Of course, all the thermodynamic properties of the field in the 
low-temperature limit are dominated by the contribution of the phonons, 
since the kink density, Eq.~(\ref{kden}), is exponentially small at low 
temperatures. However, as discussed, for example in~\cite{currie}, 
the density of kinks may be important for some transport properties that 
are insensitive to phonons, e.g. dc conductivity in charge-density-wave 
systems and transport coefficients in general \cite{rice},
which are related to the correlation functions of the field.

\section*{4. Finite-temperature thermodynamics: The transfer operator formalism}

The above approximate theory becomes inappropriate at high temperatures 
$\beta \lesssim 1$, when one can no longer neglect the interaction between the 
various nonlinear and linear excitations present in the system.
There exists an {\em exact} general formal method
for computing the grand-canonical partition function, 
by mapping the problem onto the spectral problem of a Schr\"odinger operator.  
Let us briefly apply this so-called {\em transfer operator 
formalism}~\cite{gupta,krumhansl,scalapino}
to the case of the Lam\'e potential.

The {\em canonical partition function} for the field $\phi(x)$ 
can be written as a functional integral over the
conjugated fields $\phi(x,\,t)$ and 
$\mathcal{P}(x,\,t)=\partial \phi/\partial t$:
\begin{equation}
Z_{\Lambda}(\beta)=\int \mathcal{D}\mathcal{P} \int \mathcal{D}\phi \;
\exp\left\{-\beta E_{\Lambda}[\phi,\, \mathcal{P}]\right\}\,.
\end{equation}
One can define the path integral by dividing the system into $M$ 
cells of length 
$\Delta x=2\Lambda/M$ and replacing the continuous 
fields $\phi(x,\,t)$ and $\mathcal{P}(x,\,t)$ by two discrete 
sets of $(M+1)$ field variables $\{\phi_i(t)=\phi(x_i,\,t)\}$ 
and $\{\mathcal{P}_i(t)=\mathcal{P}(x_i,\,t)\}$,
$x_i=-L+2iL/M$, $i=0,...,M$. 
At the end, one would consider
the continuum limit $M \rightarrow \infty$.
We shall consider fixed boundary conditions
\begin{equation}
\phi_0(t)=0,\, \phi_{M}(t)=2K(k)W_{\Lambda};\; \mathcal{P}_0(t)=\mathcal{P}_M(t)=0\,,
\end{equation}
that assure the constancy of both the total energy $E_{\Lambda}$ and
the topological charge $W_{\Lambda}$.  Moreover, 
we impose that $W_{\Lambda}$
{\em is an integer}, i.e., in the ``thermodynamic limit" $\Lambda \rightarrow \infty$  
the system supports topological kinks/antikinks, and  $W_{\Lambda}$ represents simply the difference
between the numbers of  kinks and  antikinks.

The partition function factorizes into a 
kinetic energy part $Z_{\Lambda}^{\mathcal{P}}(\beta)$ 
(determined by the momentum variables $\{\mathcal{P}_i\}$):
\begin{equation}
Z_{\Lambda}^{\mathcal{P}}(\beta)=\left(\frac{2\pi\,\Delta x}{\beta h^2}\right)^{(M-1)/2}\,,
\end{equation}
and a configurational (potential energy) part $Z_{\Lambda}^{\phi}(\beta)$.
The evaluation of the latter 
can be carried out using the formalism of the {\em transfer 
operator}~\cite{gupta,krumhansl,scalapino}, that allows an exact 
mapping of this problem onto the problem of finding the eigenvalues of 
an integral operator. In the continuum limit $M \gg 1$ ($\Delta x \ll 
\Lambda$), this problem can be further simplified to finding the energy 
eigenvalues of the following Schr\"odinger equation for the Lam\'e 
potential, with a temperature-dependent ``mass":
\begin{equation}
\left[-\frac{1}{2 \beta^2}\,\frac{d^2}{d\phi ^2} + 
V_L(\phi) - \varepsilon_n\right] \Phi_n (\phi)=0\,, 
\label{schrodinger}
\end{equation}
where $\Phi_n(\phi)$ are the Lam\'e 
functions~\cite{arscott,horovitz,watson}.  More precisely, 
\begin{equation}
Z_{\Lambda}^{\phi}(\beta)=\left(\frac{2 \pi \Delta x}{\beta}\right)^{M/2}\,
\sum_n \exp(-2 \Lambda \beta \varepsilon_n) \Phi_n(0) \Phi_n^*(2K(k)W_{\Lambda})\,,
\label{exp1} 
\end{equation}
where $(...)^*$ denotes the complex conjugate, and $n$ labels the ``quantum states".

In view of the periodicity of the Lam\'e potential, Eq.~(\ref{period}),
the  eigenvalues of the 
Schr\"odinger equation~(\ref{schrodinger})
lie in allowed bands, separated by forbidden bands. In each such allowed band the 
energy varies continuously with the wavenumber 
\begin{equation}
\tilde{k}\equiv \frac{q \pi}{K(k)}\,,\qquad 
\mbox {where} \quad -1/2\leqslant q \leqslant 1/2
\label{wavenumber}
\end{equation}
($\tilde{k}$ is restricted to the first Brillouin zone).
In the ``thermodynamic limit"  $\Lambda \rightarrow \infty$, and using Bloch's theorem for the eigenfunctions, one can simplify further the expression~(\ref{exp1}) to:
\begin{equation}
Z_{\Lambda}^{\phi}(\beta)= \left(\frac{2 \pi \Delta x}{\beta}\right)^{M/2}\,\int_{-1/2}^{1/2}\,dq \exp[-2
\Lambda \beta \varepsilon_1(q)-2\pi i q W_{\Lambda}] |{\Phi_q(0)}|^2 \,,
\end{equation}
where $\Phi_q(\phi)$ is the eigenfunction in Eq.~(\ref{schrodinger}) corresponding to the eigenvalue
$\varepsilon_1(q)$ that lies within the {\em first allowed band}.

The {\em total} canonical partition function in the continuum limit 
$M\gg 1$ is thus given by:
\begin{equation}
Z_{\Lambda}(\beta)=\left(\frac{2 \pi}{\beta }\right)^{M}\,\int_{-1/2}^{1/2}\,dq \exp[-2 \Lambda \beta
\varepsilon_1(q)-2\pi i q W_{\Lambda}] |{\Phi_q(0)}|^2 \,.
\end{equation}
One cannot pick out directly the most significant contribution to the integral above
as the term of largest magnitude, because the  phase $2\pi  q W_{\Lambda}$
also plays a role.  The way to avoid this problem is to go to 
the {\em grand canonical ensemble}, whose partition function is:
\begin{eqnarray}
&&\Xi_{\Lambda}(\beta,\mu)=\exp(2\Lambda \beta P)=\sum_{W_{\Lambda}=-\infty}^{\infty}
Z_{\Lambda}(\beta)\exp(\mu \beta W_{\Lambda}) =\nonumber\\
&=&\left(\frac{2 \pi }{\beta }\right)^{M}\,\int_{-1/2}^{1/2}\,dq \exp[-2 \Lambda \beta \varepsilon_1(q)] |{\Phi_q(0)}|^2 
\sum_{W_{\Lambda}=-\infty}^{\infty} \exp[-(2\pi i q - \mu
\beta)W_{\Lambda}]\,.\nonumber\\
\end{eqnarray}
Here $\mu$ is the chemical potential associated with the creation of one topological charge in the system, 
and $P$ is the thermodynamic pressure.
Recall that $W_{\Lambda}$ represents the difference between the number of topological kinks and antikinks in the system, and that is why it runs between $-\infty$ and $+\infty$.
Allowing the chemical potential $\mu$ to take imaginary values
\begin{equation}
\mu \beta=2 \pi i \lambda\,,
\end{equation}
one can immediately perform the summation over $W_{\Lambda}$, given that
\begin{equation}
\sum_{W_{\Lambda}=-\infty}^{\infty} \exp[-2\pi i (q - \lambda)W_{\Lambda}]= 
\delta(q-\lambda)\,.
\end{equation}
Thus
\begin{equation}
\Xi_{\Lambda}(\beta,\mu)=\left(\frac{2 \pi }{\beta }\right)^{M} \,
\exp[-2 \Lambda \beta \varepsilon_1(\lambda)] |{\Phi_{\lambda}(0)}|^2\,,
\end{equation}
from which one obtains the pressure
\begin{equation}
P(\mu, \beta)= \frac{M}{2\Lambda\beta} \mbox{ln}\left(\frac{2 \pi \Delta x}{\beta h}
\right) -   \varepsilon_1(\lambda)\,=
\frac{1}{\beta \Delta x} \mbox{ln}\left(\frac{2 \pi \Delta x}{\beta h}\right) -  
\varepsilon_1\left(-i\frac{\mu\beta}{2\pi}\right)\,.
\label{grandpressure}
\end{equation}
Note that the first term in the r.h.s., 
$\frac{1}{\beta \Delta x} \mbox{ln}\left(\frac{2 \pi \Delta x}{\beta h}\right)$
(corresponding to $V_L=0$), is nothing else but the contribution of the 
classical phonon field, see also~\cite{currie,krumhansl}.

We must now analytically continue the chemical potential back onto 
the real axis, which is the situation of physical relevance. 	 
This problem has been analyzed in detail in~\cite{kohn}
for general periodic symmetric potentials, 
and it was shown that under rather general circumstances
{\em the first allowed energy band of complex wavenumber maps onto
the first forbidden energy band of real wavenumber}.
 One is thus 
led finally to the important conclusion that 
in order to obtain the thermodynamics of the 
system, Eq.~(\ref{grandpressure}), 
one has to compute {\em the lowest  allowed and forbidden energy bands} 
corresponding to the associated Schr\"odinger 
equation~(\ref{schrodinger}) with the Lam\'e potential.

This is a nontrivial spectral problem, and some details are presented 
in the Appendix. 
No general results are known for arbitrary values of $\beta$ and $\nu$.
However, for the case when $2\beta^2\nu(\nu+1)$
is of the form $\ell(\ell+1)$, with $\ell$ an integer that is either
strictly positive, or $\ell \leqslant -2$, the energy bands are known 
under an implicit form that allows for combined analytical and numerical calculations.
This means that, for a fixed value of $\nu$, the thermodynamic 
properties of the field 
$\phi$  can be computed for a discrete set of 
values of the temperature
given by:
\begin{equation}
\beta=\sqrt{\displaystyle\frac{\ell(\ell+1)}{2 \nu(\nu+1)}}\;,\quad \ell \in
\mathcal{Z}\smallsetminus\{0,\,-1\}\,.
\end{equation}

As an example, we give below the analytical expression for the set of parameters
with   $2\beta^2\nu(\nu+1)=2$ (i.e., $\ell=1$). 
In this case, the expression~(\ref{grandpressure}) 
for the pressure of the system
becomes:
\begin{eqnarray}
&&P\left(\mu, \beta=1/\sqrt{\nu(\nu+1)}\right)=
\frac{M\sqrt{\nu(\nu+1)}}{2\Lambda}\; 
\mbox{ln}\left(\frac{2\pi\Delta x\sqrt{\nu(\nu+1)}}{h}\right) 
-\frac{\nu(\nu+1)}{2}\,\left[1-\frac{k'^2}{\mbox{cn}^2(\overline{\delta},\,k)}\right].  \nonumber\\
&&
\end{eqnarray}
Here $k'=\sqrt{1-k^2}$ is the complementary modulus of the elliptic functions, and 
the value of the parameter $\overline{\delta}$ follows from the implicit equation:
\begin{equation}
\frac{\mu}{2\pi\sqrt{\nu(\nu+1)}}=- \frac{\overline{\delta}}{2\,K(k')}\,-\,\frac{K(k)}{\pi}
\left[F(\gamma,\,k)-E(\gamma,\,k)+\mbox{tan}\gamma\;\,\sqrt{1-k^2\mbox{sin}^2\gamma}\,
\right]\,,
\end{equation}
where $F(\gamma,k)$ and $E(\gamma,k)$ are incomplete elliptic integrals of 
the first and second kind \cite{gradshteyn,byrd}, respectively, and 
\begin{equation}
\mbox{sin}\gamma=\mbox{sn}(\overline{\delta},\,k)\,.
\end{equation} From it one can compute, in principle, the thermodynamic properties of the system 
at this fixed temperature, e.g., the density of kinks,
\begin{equation}
n_K=\left(\frac{\partial P}{\partial \mu}\right)_T=-\frac{k'^2\sqrt{\nu(\nu+1)}}
{{2}\pi}\;\frac{\mbox{sn}(\overline{\delta},\,k)\mbox{dn}(\overline{\delta},\,k)}
{\mbox{cn}^3(\overline{\delta},\,k)}\;\left[\frac{1}{2K(k')}+\frac{K(k)}{\pi}\,
\frac{\mbox{dn}^2(\overline{\delta},\,k)}{\mbox{cn}^2(\overline{\delta},\,k)}\right]^{-1}\,.
\end{equation} 
One has now to consider the $\mu \rightarrow 0$ 
limit in the above expressions, that corresponds to the physically relevant situation of thermally 
activated excitations of the field, i.e., no external constraint on the topological charge of the system.

To complete the discussion of statistical mechanics, we need to compute 
the field-field correlation function (\ref{calx}) whose Fourier transform is 
related to the static structure factor \cite{currie,gupta,dshg}.  The 
corresponding correlation length is a measure of average distance between 
kink excitations.  As discussed in \cite{currie,deleonardis} for general periodic potentials,
$\mathcal{C}(x)$ is obtained from the knowledge of the lowest 
Lam\'e band wavefunctions \cite{horovitz,iachello}, and in the asymptotic limit $|x| \gg 1$ it is 
$\mathcal{C}(x)\sim \exp(-2n_K |x|)$, in agreement with the expression found in the 
low kink density limit, Sec.~III. Its general expression, however, cannot be obtained in a closed
analytical form.  Similarly, the correlation function involving field 
fluctuations~\cite{currie,scalapino} for the Lam\'e case remains 
an open question.  
Finally, as a general remark~\cite{currie}, the results of the transfer matrix formalism
reduce in the low-temperature limit, through a WKB-type of approximation, to the ideal kink gas 
results in Sec.~III.

\section*{5. Conclusions}
We have obtained two types of kink lattice solutions of the 
Lam\'e equation and their common limit of the single-kink solution.  
We studied the $T=0K$  ``thermodynamics"  of these kink crystals.
Using the ideal kink gas formalism, 
we studied first the approximate low-temperature thermodynamics, which 
takes into account separately the contribution of independent kinks and phonons,
while the interactions between them are not properly accounted for.  Then we 
invoked the transfer integral approach to calculate exactly the partition 
function. This maps the statistical mechanics problem onto the spectral
problem of the Schr\"odinger equation for the Lam\'e potential, see 
also~\cite{gupta}. 
Unlike the sine-Gordon equation, Lam\'e equation is not exactly solvable 
but quasi-exactly solvable: the band structure of the corresponding 
Schr\"odinger equation is  not  known analytically for all the values of the 
parameter $\ell$ of the potential.
The band structure of the Lam\'e equation is known in explicit form  
\cite{arscott,horovitz,iachello} for $\ell=1$, and in implicit form
for higher integer values of $\ell$, see the Appendix.  
Due to this  limitation, we have been able to obtain closed form expressions for 
thermodynamic quantities only for a set of temperatures.  Similar constraints 
hold in the case of another periodic  system, namely the double 
sine-Gordon equation~\cite{deleonardis}.  Although we have not been able 
to present analytic results for all temperatures, our approach provides insight into 
this system and other related periodic  quasi-exact solvable systems.

\acknowledgments
{I.B. acknowledges partial support from the Swiss National Science Foundation.
This work was supported in part by the U.S. Department of Energy.}

\appendix
\section{Spectral properties of the Lam\'e potential}
\label{lamespectrum}

Let us consider the dimensionless Schr\"odinger equation
\begin{equation}
\left[-\frac{d^2}{d\phi^2}+V_L(\phi)\right]\psi(\phi)=\varepsilon \psi(\phi)\,.
\end{equation} The Lam\'e potential with $\nu=\ell$ a strictly positive integer is well-known to be exactly solvable,
see, e.g.,~\cite{arscott,watson}. The energy  $\varepsilon$ and the wavenumber $\tilde{k}$
are expressed parametrically as
\begin{equation}
\varepsilon=\sum_{n=1}^{\ell} \displaystyle\frac{1}{\mbox{sn}^2 \alpha_n}-\left[\sum_{n=1}^{\ell}\frac{\mbox{cn}\alpha_n\,\mbox{dn}\alpha_n}{\mbox{sn}\alpha_n}\right]^2\,,
\label{appendix1}
\end{equation}
and 
\begin{equation}
\tilde{k}=i\sum_{n=1}^{\ell}Z(\alpha_n,k)-\frac{\ell \pi}{2K(k)}\,,
\label{appendix2}
\end{equation}
respectively, where $Z(u,k)$ is the Jacobi's zeta function~\cite{gradshteyn,byrd}\,,
\begin{equation}
Z(u,k)=E\left(\mbox{sin}^{-1}(\mbox{sn}(u,k)),k\right)-(E(k)/K(k))\,u\,.
\end{equation}
The parameters $\alpha_1$, ... $\alpha_{\ell}$ are subject to $(\ell-1)$ independent constraints
\begin{equation}
\sum_{p=1}^{\ell} \frac{\mbox{sn}\alpha_p\,\mbox{cn}\alpha_p\,\mbox{dn}\alpha_p+
\mbox{sn}\alpha_n\,\mbox{cn}\alpha_n\,\mbox{dn}\alpha_n}{\mbox{sn}^2\alpha_p - \mbox{sn}^2\alpha_n} =0, \quad p\neq n,\quad n=1,...,(\ell-1)\,.
\label{appendix3}
\end{equation}
The allowed energy bands correspond to a real value of the wavenumber $\tilde{k}$, i.e., to the condition
\begin{equation}
\mbox{Re}\left[\sum_{n=1}^{\ell}Z(\alpha_n,k)\right]=0\,, 
\label{appendix4}
\end{equation}
and there are $(\ell+1)$ allowed bands ($\ell$ finite bands followed by a continuum band), separated by $\ell$ forbidden bands. 

Closed analytical results can be obtained in the $\ell=1$ case. In particular, for the lowest allowed  band one obtains the parametric 
equations~\cite{iachello,watson} for the energy 
\begin{equation}
\varepsilon_1=1-k'^2\mbox{cn}^2(\eta+K(k'),k')\,, \quad k^2\leqslant \varepsilon_1 \leqslant 1\,,\quad 0\leqslant \eta \leqslant K(k')\,,
\end{equation}
and the corresponding wavenumber
(for half of the first Brillouin zone, the other half being obtained by symmetry)
\begin{equation}
\tilde{k}=Z(\eta,k')-\frac{\pi}{2K(k)}\left(1-\frac{\eta}{K(k')}\right)-\sqrt{\frac{(\varepsilon_1 -k^2)(1-\varepsilon_1)}{1+k^2-\varepsilon_1}}\,.
\end{equation}
Here $k'=\sqrt{1-k^2}$ is the complementary modulus of the elliptic functions. With a change of 
parametrization and using the properties of the elliptic 
functions~\cite{gradshteyn,byrd}, these formulae can be reduced to the following (more familiar to physicists) 
expressions~\cite{horovitz,shastry}:
\begin{equation}
\varepsilon_1 =1-k'^2\mbox{cn}^2(p,k')\,, \quad k^2\leqslant \varepsilon_1 \leqslant 1\,,\quad -K(k')\leqslant p \leqslant K(k')\,,
\label{allowed1}
\end{equation}
and for the wavenumber in the first Brillouin zone
\begin{equation}
\tilde{k}=Z(p,k')+\frac{\pi p}{2K(k)K(k')}\,.
\label{kallowed1}
\end{equation}

Note that (as demonstrated in a more general context in~\cite{kohn})
the results for the first forbidden band can 
be obtained {\em formally} from those
concerning the first allowed band simply by replacing 
$p$ in~(\ref{allowed1}) and~(\ref{kallowed1}) by
$i\overline{p}$. 
One obtains thus for the energy levels of the first forbidden band:
\begin{equation}
\overline{\varepsilon}_1=1-\frac{k'^2}{\mbox{cn}^2(\overline{p},\,k)}\,,
\end{equation}
where the parameter $\overline{p}$ ($-K(k')\leqslant \overline{p} \leqslant K(k')$)
corresponds to a {\em complex}
wavenumber given by:
\begin{equation}
\tilde{k}=i\,
\left[F(\gamma,\,k)-E(\gamma,\,k)+\mbox{tan}
\gamma\;\sqrt{1-k^2\mbox{sin}^2\gamma}\right]\,+ 
i \frac{\overline{p}\pi}{2K(k)K(k')}\,,
\end{equation}
with 
\begin{equation}
\mbox{sin}\gamma=\mbox{sn}(\overline{p},\,k)\,.
\end{equation}

One realizes, however, that with increasing $\ell$ the 
complexity of the implicit equations (\ref{appendix1}),
(\ref{appendix2}),(\ref{appendix3}), and (\ref{appendix4})
(that one has to solve in order to obtain the dispersion 
relations for the allowed bands) increases, up to becoming 
prohibitive even for present-day powerful computers.

Fortunately, a recent paper by Maier~\cite{maier03} offers an 
alternative to the above-described Hermite-Halperin method
of band computation.
It is shown, using the so-called Hermite--Krishever Ansatz,
that the dispersion relations for any integer $\ell$
can be expressed in terms of the $\ell=1$ relations. Moreover, 
the effective construction of these dispersion relations,
whose technical details will not be given here, makes use of 
relatively simple polynomials and rational functions.
This represents an enormous gain in the efficacy of
symbolic calculation of the dispersion relations.

To our knowledge, there is no information yet
on the spectrum of the Lam\'e potential when $\nu$ is not a positive integer. 
In these cases one has to obtain the results strictly from numerical 
evaluations.

\newpage
\begin{center}
{\bf FIGURE CAPTIONS}
\end{center}

\noindent FIGURE 1: Typical profile of the Lam\'e potential $V_L(\phi)$
over a period $[0,\,2K(k)]$.\\

\noindent FIGURE 2: Typical profile of a kink lattice solution of the first type
(an array of nontopological kink-antikink pairs) 
over a spatial period $4L$ of the lattice 
(for a fixed value of the control parameter $a^2$, $0< a^2 <1$.)\\

\noindent FIGURE 3: The stability diagram of the kink 
lattice solution of the first type, at $T=0K$,
in the plane of the parameters $k^2$ (of the Lam\'e potential) 
and $a^2$ (the control parameter).\\

\noindent FIGURE 4: Typical profile of a kink lattice solution of the second type
(an array of topological kinks/antikinks) 
over a spatial period $4L$ of the lattice 
(for a fixed value of the control parameter $a^2>0$).
Note that we have 
plotted $|\tilde\phi(x)|$ modulo $2K(k)$; there
are no ``cusps" in the field $\tilde\phi(x)$. \\

\noindent FIGURE 5: The profile of the topological single kink solution.

\newpage

\begin{center}
{\bf FIGURE 1}
\end{center}

\newpage 

\begin{center}
{\bf FIGURE 2}
\end{center}

\newpage

\begin{center}
{\bf FIGURE 3}
\end{center}

\newpage

\begin{center}
{\bf FIGURE 4}
\end{center}

\newpage

\begin{center}
{\bf FIGURE 5}
\end{center}

\end{document}